\documentclass[aps,floatfix]{revtex4}

\usepackage{graphicx}%
\usepackage{dcolumn,float}%
\newcommand{\no}{\noindent}

\newcommand{\beq}{\begin{equation}}
\newcommand{\eeq}{\end{equation}}
\newcommand{\qq}{\begin{eqnarray}}
\newcommand{\qqq}{\end{eqnarray}}
\newcommand{\ve}{\varepsilon}
\newcommand{\h}{{\cal H}_{N,M}}

\begin{document}

\title{The elementary excitations
of the BCS model in the canonical ensemble}

\date{\today}

\author{Germ\'an Sierra,$^{1}$\footnote{Talk presented
at the 6th International workshop 
on Conformal Field Theory and Integrable Models,
Chernologka, Russia, Sept.~2002.} 
Jos\'e Mar\'{\i}a Rom\'an$^{1}$
and  Jorge\ Dukelsky$^{2}$ } 

\affiliation{ 
$^{1}$Instituto de F\'{\i}sica Te\'orica, CSIC/UAM,
Madrid, Spain.\\
$^{2}$Instituto de Estructura de la Materia, CSIC, Madrid, Spain
}

\begin{abstract}
We summarize previous works on the exact ground state
and the elementary excitations of the 
exactly solvable BCS model in the canonical ensemble.
The BCS model is solved by Richardson equations, and, 
in the large coupling limit, by Gaudin equations.
The relationship between this two kinds of solutions are
used to classiffy the excitations.
\end{abstract}


\maketitle

\newcommand{\bb}{\boldsymbol{\beta}}
\newcommand{\ba}{\boldsymbol{\alpha}}

\section{Introduction}

One of the most fundamental problems in a many body
system is to know which are the ground state (GS) 
and the low energy excited states, which determine the  
thermodynamic properties and the response to
external fields. In most cases the excited states
can be understood in terms of a collection
of elementary excitations characterized by
their statistics, discrete quantum numbers
as spin, charge and dispersion relation. 
In the pairing model
of superconductivity, proposed by  Bardeen, Cooper and Schrieffer
in 1957, this problem was solved long ago
in the grand canonical ensemble for a large number of particles 
\cite{BCS,T}. 
However recent studies, motivated by the fabrication of ultrasmall
metallic grains, show that the grand canonical
BCS solution deviates strongly from the exact 
numerical or analytical solution for systems 
with a fixed and small number of particles 
(for a review see \cite{vDR}). 
Most of the previous studies have focused on the ground
state of the BCS system, and some of its
excitations. In this contribution we shall 
review further progress concerning the
understanding of the full excitation
spectrum \cite{rsd}.

\section{The grand canonical BCS ansatz}

The BCS model of superconductivity
is characterized by the energy levels
of the electrons and the scattering potential
between them. When the latter is a constant, 
the BCS model is exactly solvable \`a la Bethe
\cite{exact}
and integrable \cite{CRS}. This is the model we shall 
used to study the GS and excitations. 
The Hamiltonian of the reduced BCS is defined by 
\cite{vDR}
\beq
H_{BCS} = \frac{1}{2}\sum_{j, \sigma= \pm} 
\ve_{j\sigma} c_{j \sigma}^\dagger c_{j \sigma}
  - G \sum_{j, j'}  c_{j +}^\dagger c_{j -}^\dagger 
c_{j' -} c_{j' +} \; , 
\label{1}
\eeq
where $c_{j,\pm}$ (resp.\ $c^\dagger_{j,\pm}$)
is an electron  
annihilation (resp.\ creation) operator   
in the time-reversed states $|j, \pm \rangle$
with energies $\varepsilon_j/2$, and
$G$ is the BCS dimensionful coupling constant.
The sums in (\ref{1}) run over a set of $N$ doubly
degenerate energy levels $\varepsilon_j/2 \; (j=1,\dots, N)$.
We adopt Gaudin's notation, according to which
$\varepsilon_j$ denotes the energy of a 
pair occupying the level $j$ \cite{G-book}. 
To fix ideas we shall use 
the so called equally space or picket fence
model, that is the one employed in the study of 
ultrasmall superconducting grains \cite{vDR}. 
This model is given by the choice
$\ve_j = d (2 j - N -1)$,
where $d=\omega/N$ is the single particle energy 
level spacing and $\omega/2$ is the Debye energy.
The coupling $G$ can be written as  
$G= g d $, where $g$ is dimensionless.
 
The BCS ansatz for the GS of this Hamiltonian
in the grand canonical ensemble is given by \cite{BCS,T}
\beq
|BCS \rangle = \prod_j (u_j + v_j \; c^\dagger_{j,+}
\, c^\dagger_{j,-}) \; |0 \rangle ,
\label{2}
\eeq
where $|0 \rangle$ is the Fock vacuum of the electron
operators and $u_j, v_j$ are the BCS variational
parameters given by 
\begin{eqnarray}
u_j^2 = \frac{1}{2} \left( 1 + \frac{\xi_j}{E_j} \right),
& \qquad &
v_j^2 = \frac{1}{2} \left( 1 - \frac{\xi_j}{E_j} \right),
\label{3} \\
\xi_j = \ve_j - \ve_0 - G , 
& \qquad &
E_j = \sqrt{\xi_j^2 + \Delta^2} . 
\label{4}
\end{eqnarray}
In these eqs. $\ve_0$ and $\Delta$ are twice the chemical 
potential and the BCS gap, which are found by solving
the following equations:
\beq
\frac{1}{G} = \sum_j \frac{1}{E_j} , \qquad
M = \frac{1}{2} \sum_j \left( 1 - \frac{\xi_j}{E_j} \right),
\label{5} 
\eeq
where $M$ is the number of electron pairs. 
 The most studied case
in the literature corresponds to the half-filled situation,  
where the number of electrons,  $N_e= 2M$, equals the number of levels
$N$ \cite{vDR}. In this case the solution of eqs.~(\ref{5})
in the large $N$ limit is given by  
 $\Delta = \omega/\sinh(1/g)$ and $\ve_0=0$.

The excited states in the grand canonical ensemble can be obtained
acting on the GS ansatz (\ref{2}) with the Bogoliubov
operators $\gamma_{j,\sigma} \; (\sigma = \pm)$ 
\beq
 \gamma_{j_1,\sigma_1} \dots 
\gamma_{j_n,\sigma_n} |BCS \rangle , 
\qquad \gamma_{j, \pm} = u_j \, c_{j \pm} 
\mp v_j \, c^\dagger_{j \mp},
\label{6} 
\eeq
and have an energy $\frac{1}{2} (E_{j_1} + \dots + E_{j_2})$.
Recall that in our conventions $E_j$ is twice
the quasiparticle energy, thus the factor $1/2$ in the
energy of the states (\ref{6}).

\section{Pair-hole representation of BCS model}

Every energy level $j =1, \dots, N$ has four
possible states given by
\begin{eqnarray}
|0 \rangle  : & & 
\mbox{empty,}  \nonumber \\
c^\dagger_{j \sigma} |0 \rangle :  & & 
\mbox{singly occupied,} \label{7} \\
c^\dagger_{j +} c^\dagger_{j -} |0 \rangle : & &
\mbox{doubly occupied.} \nonumber
\end{eqnarray}
An important property of the Hamiltonian
(\ref{1}) is the ``blocking'' of 
levels which are singly occupied, which 
means that these levels decouple from the rest of the system. 
This is a consequence of the equation
\beq
H_{BCS} \; c^\dagger_{j \sigma} |\psi \rangle
= \frac{\ve_j}{2} \;  c^\dagger_{j \sigma} \;  |\psi \rangle
+  c^\dagger_{j \sigma}\;  H_{BCS} |\psi \rangle ,
\label{8}
\eeq
where the state $|\psi \rangle$ does not contain
the operators $c^\dagger_{j \sigma}$. Thus 
the singly occupied levels 
only contribute with their
kinetic energy and one can study the dynamics
on those levels which are either empty
or doubly occupied.

Let us call $\h$ the Hilbert space of states 
of $M$ pairs distributed among $N$
different energy levels. To describe 
$\h$ let us define 
the hard-core boson operators 
\beq
b_j = c_{j,-} c_{j,+}, \qquad 
b_j^\dagger= c^\dagger_{j,+} c^\dagger_{j,-} , \qquad 
N_j = b^\dagger_j b_j ,
\label{9}
\eeq
which satisfy the commutation relations
\beq
[ b_j, b_{j'}^\dagger ] = \delta_{j,j'} \; (1 - 2 N_j). 
\label{10}
\eeq
The Hamiltonian (\ref{1}) restricted to the
Hilbert space $\h$ 
can then be written as
\beq
H_{BCS} = \sum_{j}  \varepsilon_j b^\dagger_j b_j - G \,
\sum_{j,j'} \; b_j^\dagger b_{j^{\prime}} \; .
\label{11}
\eeq
A basis of states of $\h$ is given by 
\beq 
|I \rangle = \prod_{j \in I} b^\dagger_j \; | 0 \rangle,
\label{12}
\eeq
where $I$ denotes a set of $M$ different integers
ranging from $1$ to $N$. Thus the dimension
of $\h$ is given by the combinatorial number
$C^N_M$. A convenient pictorial representation 
of the singly and occupied states 
is given by \cite{rsd}
\beq
\circ \leftrightarrow |0 \rangle , \qquad
\bullet  \leftrightarrow  b^\dagger_j \, |0 \rangle .
\label{13}
\eeq
At $G=0$ the ground state of (\ref{11})
is the Fermi sea obtained filling all the levels
below the Fermi energy $\ve_F$. The set $I_0$
is given in this case by $I_0 = \{ 1,2,\dots,M \}$
(see fig.~1). Similarly  the lowest energy  
excited state at $G=0$ corresponds to the choice
 $I_1 = \{ 1,2,\dots,M-1, M+1 \}$ (see fig.~2). 
\begin{figure}[t]
\begin{center}
\includegraphics[width= 10 cm,angle= 0]{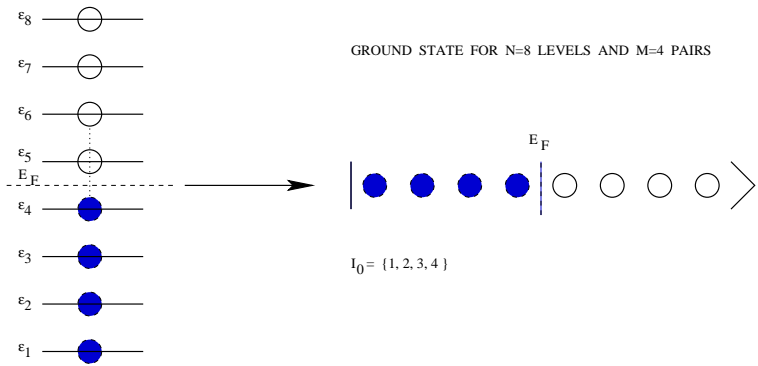}
\end{center}
\caption{Pair-hole representation of the ground
state at $G=0$ for $N=2M=8$.  
}
\label{fig1}
\end{figure}
\begin{figure}[ht]
\begin{center}
\includegraphics[width= 10 cm,angle= 0]{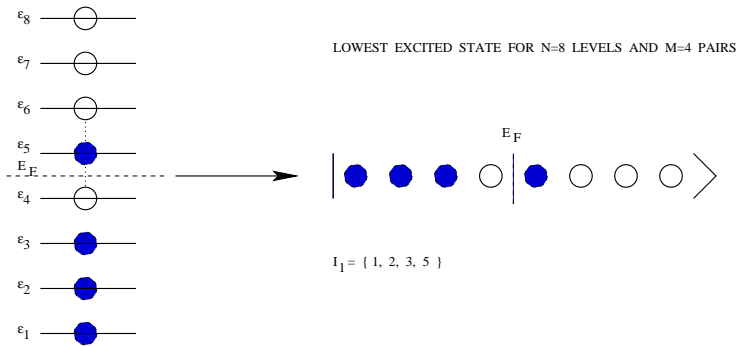}
\end{center}
\caption{Pair-hole representation of the lowest
energy excited state at $G=0$ for $N=2M=8$.  
}
\label{fig2}
\end{figure}

\section{Exact solution of the  BCS model in the canonical ensemble}

In 1963 Richardson showed that the eigenstates of the 
Hamiltonian (\ref{11}) 
with $M$ pairs have the (unnormalized) product form 
\cite{exact} 
\beq
|M \rangle = \prod_{\nu = 1}^M B_\nu^{\dagger} |{\rm vac} \rangle, 
\qquad
B_\nu^{\dagger} = \sum_{j=1}^N \frac{1}{\varepsilon_j - E_\nu} 
\; b^\dagger_j \; ,
\label{14} 
\eeq
where the  parameters $E_\nu$ ($\nu = 1, \dots , M$) are, 
in general, complex solutions of the $M$ coupled algebraic
equations 
\beq
\frac{1}{G } 
= \sum_{j=1}^N \frac{1}{ \varepsilon_j - E_\nu}
- \sum_{\mu=1 (\neq \nu)}^{M} \frac{2}{ E_\mu - E_\nu}\;, 
\qquad \nu =1, \dots, M ,
\label{15}
\eeq
which play the role of Bethe ansatz equations for this
problem \cite{B,BF,S,AFF,ZLMG,vDP,Scomo,AFS}. 
The energy of these states is given by the sum of the
auxiliary parameters $E_\nu$, i.e.\
\beq
{E} (M) = \sum_{\nu = 1}^{M} E_\nu.
\label{16}
\eeq
The ground state  of $H_{BCS}$ is given by the solution
of eqs.~(\ref{15}) which gives the lowest value of ${E}(M)$. 
The (normalized) states (\ref{14}) can also be written  
as \cite{R-norm}
\beq
|M \rangle = \frac{C}{ \sqrt{M!} } \sum_{j_1, \cdots , j_{M}}
\psi(j_1, \dots,  j_{M}) b^\dagger_{j_1} \cdots b^\dagger_{j_{M}}
|{\rm vac} \rangle, 
\label{17}
\eeq
where 
the sum excludes double occupancy of pair states
and the wave function $\psi$ takes the form
\beq
\psi(j_1, \cdots, j_{M}) = \sum_{\cal P}
 \prod_{k=1}^{M} \frac{1}{ \varepsilon_{j_k} 
- E_{{\cal P}k} }\; .  
\label{18} 
\eeq
The  sum in (\ref{18})  runs  over all the 
permutations, ${\cal P}$, 
of $1, \cdots, M$. The constant  $C$ in (\ref{17})
guarantees the normalization of the state \cite{R-norm} (i.e.\
$\langle M|M\rangle =1$). 

The number of solutions of the Richardson's eqs.~(\ref{15}) 
is equal to the dimension of the Hilbert
space $\h$, namely $C^N_M$ \cite{G-book}. For finite and small
values of $N$, $M$ the solutions $\{ E_\mu(G) \}_{\mu =1}^M$ 
have to be found numerically. These solutions and
the corresponding eigenstates  can be classified
according to the values taken by $E_\mu(G)$ 
at $G=0$, namely \cite{R-roots}
\beq
\lim_{G \rightarrow 0} E_\mu (G) = \ve_j , 
\;\; j \in I ,
\label{19}
\eeq
where $I$ is the set $I$ that label the basis
(\ref{12}) of $\h$. This means that the spectrum
of the BCS Hamiltonian follows an adiabatic
evolution as a function of $G$ \cite{R-roots,large}
For very small values of $G$ all the 
roots $E_\mu(G)$ 
of eq.~(\ref{15}) are real and close
to their $G=0$ value given in eq.~(\ref{19}).
This happens for all the states labelled by $I$.

\subsection*{Ground State solution}

\begin{figure}[t!]
\begin{center}
\includegraphics[width= 9 cm]{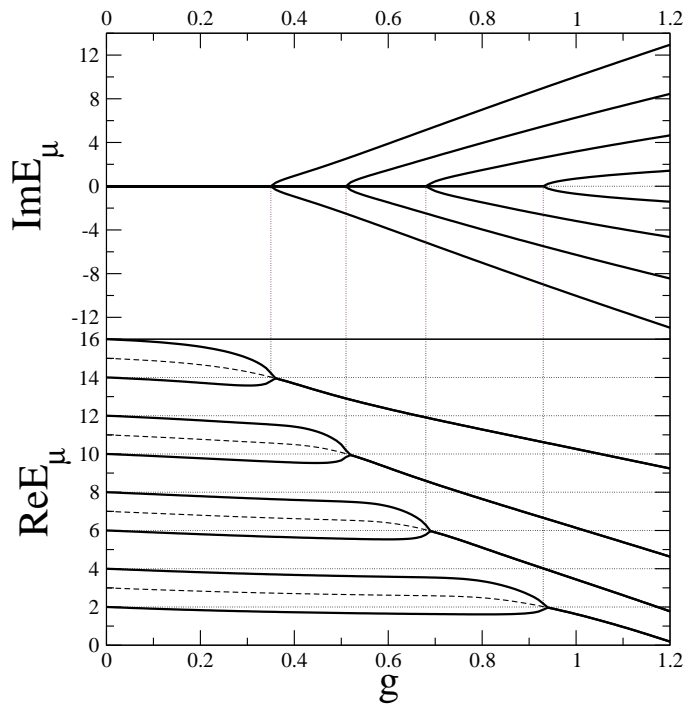}
\end{center}
\caption{Evolution of the real and imaginary parts
of $E_\mu(g)$, in units of $d = \omega/N$, 
for the equally spaced model with $M=N/2=8$, 
as a function of the coupling
constant $g = G/d$ 
\cite{R-roots,large}. For convenience the energy levels
are choosen in this figure as $\ve_j = 2 j$.
}
\label{fig3}
\end{figure}

In particular for the GS, i.e.\ $I_0$, as we increase
$G$ the real roots $E_M$ and $E_{M-1}$, nearest to the
Fermi level, approach from above and below the energy
$\ve_{M-1}$, and become equal to it at some critical
value $G= G_{c1}$. For $G > G_{c1}$ the two roots
$E_M$ and $E_{M-1}$ become a complex conjugate pair.
Increasing $G$ further one encounters
a similar phenomena for the  roots
$E_{M-2}$ and $E_{M-3}$, and so on, 
until all the roots become complex
(see fig.~\ref{fig3}).

In the limit where $N$ is large, while 
$M/N$ and $\omega = d N$ remain finite 
the complex roots $E_\mu$ form an 
open arc $\Gamma$ 
with end points $\ve_0 \pm i \; \Delta$
(see fig.~4) 
\cite{G-book,R-limit,large}. 
This gives an interesting geometrical
meaning to the chemical potential
($\ve_0/2$) and the BCS gap ($\Delta/2$)
for the exactly solvable BCS model.
There are also roots which stay real
staying in the segment 
$(- \omega, \ve_1)$ where $\ve_1$
is the intersection point 
of $\Gamma$ with the real axis.

\begin{figure}[t!]
\begin{center}
\includegraphics[height= 10 cm,angle= -90]{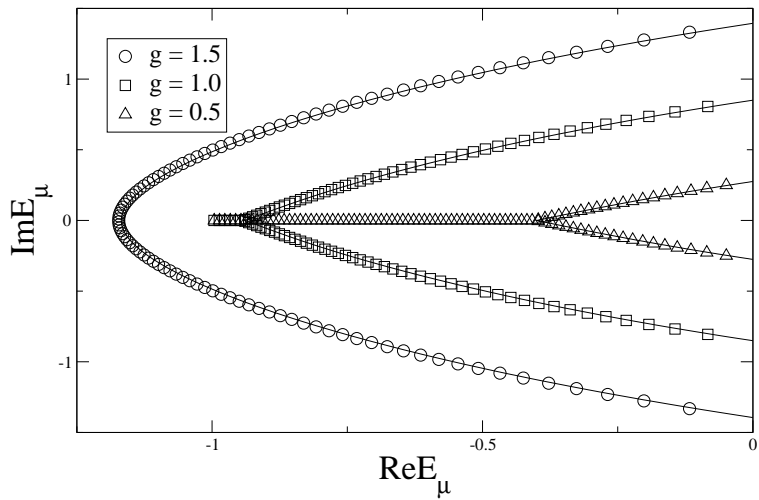}
\end{center}
\caption{Plot of the roots $E_\mu$ for the equally spaced model
in the complex plane \cite{large}.  
The discrete symbols denote the numerical values for $M=100$. 
The continuous lines are the analytical curves
obtained in \cite{large}. All the energies are in units
of $\omega$.
}
\label{fig4}
\end{figure}

\section{Excitations of the canonical BCS model}

In the canonical ensemble, where
the number of electrons is fixed, 
the excited states 
can be obtained from the GS in two ways: 
by {\it breaking Cooper pairs} or
by {\it pair-hole excitations} \cite{rsd}.

\subsection*{Breaking Cooper pairs}

In fig.~5 
we show at $G=0$ the excited state obtained
by breaking the electron pair nearest to the Fermi level.
The spin up electron remains in the same energy level,
while the spin down goes one level up, 
producing two blocked levels. As shown in
fig.~5 the problem becomes identical to that
of two decoupled levels and a system with 
two less levels active for pairing interactions. 
Hence for $G >0$ the energy of this excitation
is the sum of the single particle 
energy of the blocked levels
plus the GS energy of the system with these
levels being removed. This situation
is similar to what happens when the system
has an odd number of electrons or when
it is under the effect of an external magnetic
field \cite{vDR}. In both cases there are
singly occupied levels, which only
contribute with their non interacting energy. 
This sort of excitations are the ones that
have been mainly studied in the literature.

\begin{figure}[h!]
\begin{center}
\includegraphics[height= 5 cm]{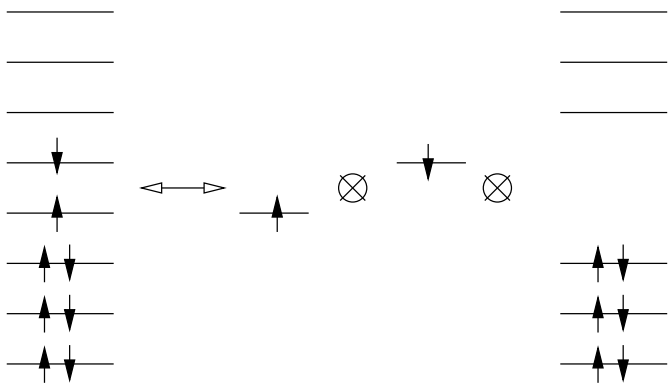}
\end{center}
\caption{Excited state obtained by breaking 
a Cooper pair at $G=0$. On the rhs the  
singly occupied levels are blocked and decouple
from the rest of the system.
}
\label{fig5}
\end{figure}

\subsection*{Pair-hole excitations}

The pair-hole excitations 
consist in promoting a pair
below the Fermi level to a level above
it. At $G=0$ the resulting state 
can be viewed as a pair-hole excitation
of the Fermi sea (see fig.~6). 
This sort of excitations belong
to the Hilbert space $\h$ and were
first consider in reference \cite{rsd}. 
It is clear that a general excitation
consists in the combination
of broken Cooper pairs and 
pair hole-excitations. 
The rest of this review will focuse
on the latter ones.

\begin{figure}[h!]
\begin{center}
\includegraphics[height= 5 cm]{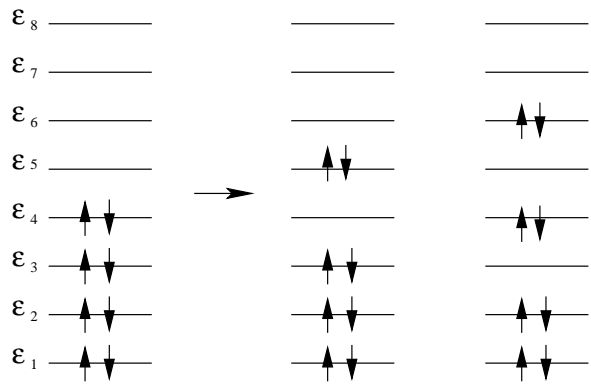}
\end{center}
\caption{Two possible 
excited states obtained by moving one and
two pairs above the Fermi level. 
}
\label{fig6}
\end{figure}

In fig.~7 we plot the real part of $\{ E_\mu(g) \}_{\mu=1}^M$
for three states, labelled $I_0, I_1, I_2$ and 
a system with $N=40$ energy levels and $M=20$ pairs
as a function of the BCS dimensionless coupling $g$ from
0 to 1.5. The states are given by the choices
\beq
I_0 =  \{ 1,\dots, 18,19,20 \}, \quad
I_1 =  \{ 1,\dots, 18,19,21 \}, \quad
I_2 =  \{ 1,\dots, 18,20,22 \}. \label{20} 
\eeq
The state $I_0$ is the GS of the system and the
pattern that follows ${\rm Re} E_\mu(g)$ 
is the same as in fig.~3. Notice that as
$g \rightarrow \infty$ all the roots
become complex and escape to infinity. 
The state $I_1$ is the lowest energy state
of the system, and as we see in fig.~7, 
in the limit $g \rightarrow \infty$, 
the root $E_{20}(g)$ stays real and finite
while the remaining $M-1=19$ roots
escape to $\infty$. Finally, the state $I_2$
has two roots  $E_{20}(g)$ and 
$E_{19}(g)$ that stay real and finite 
in the $g \rightarrow \infty$ limit. 
This is a general feature of the numerical
solutions of the Richardson's equations
(\ref{15}), namely in the limit 
 $g \rightarrow \infty$ there are
$N_G$ roots $E_\mu(g)$ 
that remain finite, not necessarily real,
while the $M - N_G$ remaining ones go to $\infty$. 
As we shall see below the number $N_G$
of finite roots can be interpreted 
as the number of elementary excitations
of a given excited state.

\begin{figure}[h!]
\begin{center}
\includegraphics[width= 10 cm]{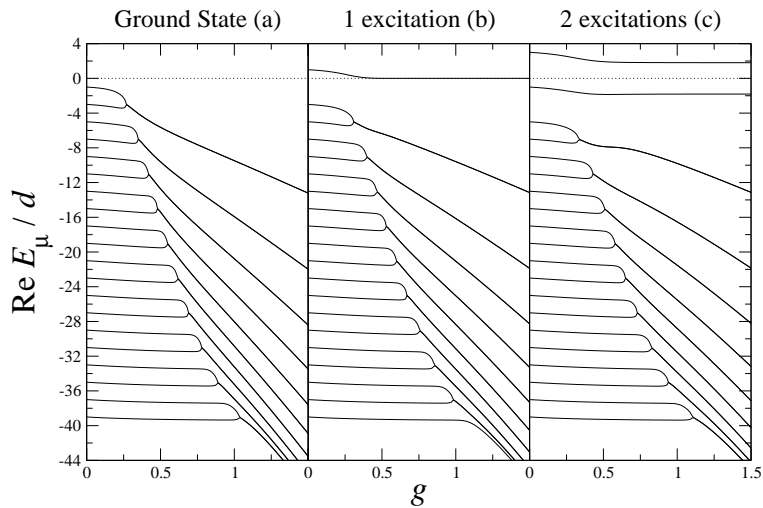}
\end{center}
\caption{Real part of  $E_\mu$ 
for the equally spaced model 
with $M=N/2=20$ pairs 
and $N_G =0,1,2$ excitations \cite{rsd}.}
\label{fig7}
\end{figure}

\subsection*{Excitation energies}

\begin{figure}[t!]
\begin{center}
\includegraphics[width= 7 cm]{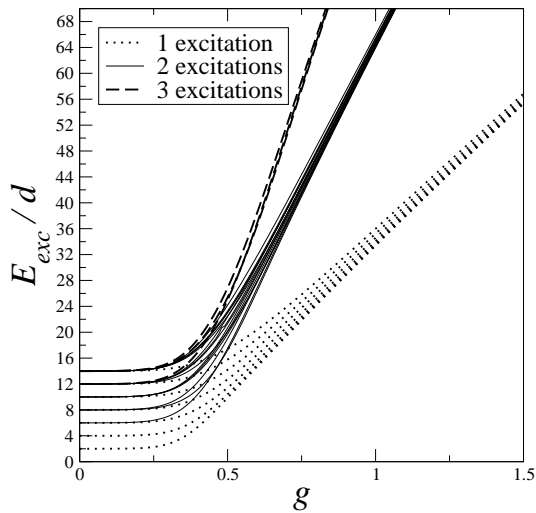}
\end{center}
\caption{Excitation energies
$E_{exc}=E - E_{GS} \leq 14 d$ 
for $M= 20$ pairs at half filling \cite{rsd}.
There are 44 =13+26+5 states 
corresponding to $N_G= 1,2,3$
respectively. The particle-hole symmetry
reduces these numbers
to 25 = 7+15+3. }
\label{fig8}
\end{figure}

To support this conjecture we have computed
the excitation energy $E_{\rm exc} = E - E_{GS}$
of some low lying 
excited states. Fig.8 shows the energies 
$E_{\rm exc}$ for a system with $M= N/2=20$
as a function of $g$. At $g=0$ 
$E_{\rm exc}$ is simply given by the energy
needed to lift the pairs from the Fermi sea up 
to the  unoccupied energy levels,
but as $g$ increases the excitation energies
quickly converge to asymptotas whose
slope is given essentially by $N_G$,
\beq
\lim_{g \rightarrow \infty} \; E_{\rm exc}
= N_G \, \Delta , \qquad \Delta \sim g \omega . 
\label{21}
\eeq
Moreover using the methods of Refs.~\onlinecite{G-book,large} 
one can show in the large $N$ limit that  
the excitation energy of a Richardson state  is given by
\beq
E_{exc} = \sum_{\alpha =1}^{N_G} 
\sqrt{ (E_\alpha - \ve_0)^2 + \Delta^2} \;,
\label{22}
\eeq
where $\ve_0$ is twice the chemical potential,
and the energies $\{ E_\alpha \}_{\alpha=1}^{N_G}$ 
are the ones that remain finite in the $g \rightarrow \infty$
limit.  In the latter limit 
one has $\Delta \sim g \omega$ and 
eq.~(\ref{22}) becomes eq.~(\ref{21}). 
The excitation energy given by eq.~(\ref{22}) 
fits quite well the excitation energies of our prototype
example ($N = 40$, $M = 20$), as shown in fig.~\ref{fig8}.

In order to compare our results with the BCS standard solution
let us consider the excitation energy of a {\em real Cooper
pair}, $\gamma^\dagger_{j+} \gamma^\dagger_{j-}$ 
in the Bogoliubov approach, which is given by
$\sqrt{\ve_j^2 + \Delta^2}$ (notice that $\Delta \equiv 2\Delta_{BCS}$).
The standard Bogoliubov quasiparticle with an energy 
$\frac{1}{2} \sqrt{\ve_j^2 + \Delta^2}$ would have to be compared 
with excitations involving broken Cooper pairs.
Since $E_{\alpha}$ in eq.~(\ref{22}) lies between two energy levels, with
$\ve_{j+1} - \ve_j = 2d \sim 1/N$ (e.g.\ in fig.~\ref{fig7}b 
$E_{20}(\infty) = 0$ with $\ve_{20} < E_{20} < \ve_{21}$),
$E_{\alpha} = \ve_j + O(1/N)$.
Therefore, our theory is consistent within $O(1/N)$ 
corrections, as it is well known from the existing relation 
between the canonical and  grand canonical ensembles.

\section{Counting the number of states with $N_G$ elementary excitations}

We said above that the dimension
of the Hilbert space $\h$ of the states
with $M$ pairs distributed into $N$ different 
energy levels is given by the combinatorial
number $C^N_M$, which is also equal
to the number of solutions of the
Richardson's equations (\ref{15}). 
The next question is: how many of these
solutions contain $N_G$ roots
that remain finite in the $g \rightarrow \infty$
limit? Let us call $d_{N,M,N_G}$ that number
which must obviously satisfy 
$C^N_M = \sum_{N_G=0}^M \; d_{N,M,N_G}$. 
In reference \cite{G-book} Gaudin made
the following conjecture 
\beq
d_{N,M,N_G} = \left\{ 
\begin{array}{ll}
C^N_{N_G} - C^N_{N_G -1} & 0 \leq N_G \leq M \\
0 & N_G > M 
\end{array}
\right. .
\label{23}
\eeq
In table 1 we give some examples of this formula.
Since $d_0=1$,  there is only one state where all
the energies $E_\mu$ go to infinity as
$g \rightarrow \infty$, which is nothing but
the GS of the system.
\begin{center}
\begin{tabular}{|c|c|c|c|c|c|}
\hline
N$^o$ of solutions & $d_M$ & $d_{M-1}$ & $\cdots$ & $d_1=N-1$ & $d_0=1$ \\
\hline
$E_\mu$ finite & $M$ & $M-1$ & $\cdots$ & 1 & 0   \\
$E_\mu$ infinite & $0$ & $1$ & $\cdots$ & $M-1$ & $M$ \\   
\hline
\end{tabular}

\vspace{0.3 cm}
Table 1.- Classification of roots in the $g \rightarrow \infty$ limit.
Here $d_{N_G}$ stands for $d_{N,M,N_G}$. 
\end{center}
This conjecture can be motivated as follows.
Suppose that in the limit $g \rightarrow \infty$
there is a set $\{ E_\alpha(\infty)\}_{\alpha=1}^{N_G}$
of $N_G$ roots that remain finite while the remaining
$M - N_G$ ones escape to infinity. Then from eqs.~(\ref{15})
one derives that 
the roots $E_\alpha = E_\alpha(\infty)$ satisfy  
\beq
0 
= \sum_{j=1}^N \frac{1}{ \varepsilon_j - E_\alpha}
- \sum_{\beta \neq \alpha}^{N_G} \frac{2}{ E_\beta - E_\alpha}\;, 
\qquad 
\alpha =1, \dots, N_G. 
\label{24}
\eeq
\no 
These equations are due to Gaudin and they appear
in the diagonalization of a spin chain model known as Gaudin 
magnets \cite{Gaudin}.
The number of solutions of eqs.~(\ref{24}) was shown by Gaudin
to be given by (\ref{23}) \cite{G-book}. 
In reference \cite{rsd}
we checked numerically this Gaudin's conjecture for
small systems. Furthermore we were able to show,
in the large $N$ limit, that for finite 
$g$ the roots $E_\alpha= E_\alpha(g)$ satisfy
an ``effective'' Gaudin equations
\beq 
0= \sum_{j=1}^N \frac{1}{ R(\ve_j) (\ve_j - E_\alpha)}
- \sum_{\beta \neq \alpha}^{N_G} 
\frac{2}{ R(E_\beta)( E_\beta - E_\alpha)} \;, 
\label{25}
\eeq
with $R(E) = \sqrt{ (E - \ve_0)^2 + \Delta^2}$.
As $g \to \infty$ one has $\Delta \sim g \omega$ and 
eqs.~(\ref{25}) become eqs.~(\ref{24}).

\section{A classification problem}

Gaudin's conjecture (\ref{23}) gives the number 
$d_{N,M,N_G}$ of states with a given value of $N_G$, 
but says nothing of how to find them.
Recall that the solutions of Richardson's eqs.~(\ref{15}) 
are obtained by starting from a given initial
state at $g=0$, labelled by the set $I$, 
and then increasing the value of $g$. Hence
the problem is to find for each state $I$ which is 
the number of roots $N_G$, which remains finite
in the $g \rightarrow \infty$ limit. 
This defines the function $N_G(I)$ which,
in physical terms, gives the number of elementary
excitations present in the state. The total
number of states $I$ with $N_G = N_G(I)$
is nothing but $d_{N,M,N_G}$. 

Finding $N_G(I)$  is a highly non trivial problem
because it connects the two extreme cases 
$g=0$ and $g=\infty$. From a numerical point
of view it is also a challenging problem due to the
existence of several sorts of singularities in the 
merging and splitting of roots with no obvious
a priori pattern, except for the GS. Despite of these
difficulties it is possible to give a simple
algorithm yielding  $N_G(I)$, which has an interesting 
combinatorial interpretation in terms of 
Young diagrams \cite{rsd}.
Before we give this algorithm we need some 
preliminary definitions.

\subsection*{Path representation of states and Young diagrams}

The first idea is to associate to every state $I$, 
in the Hilbert space $\h$, a path $\gamma_I$ in the lattice
${\cal Z}^2$. This is done, in the pair-hole
representation of $I$, by associating  to every empty level,
$\circ$, 
a horizontal link in ${\cal Z}^2$, and to every
occupied level, $\bullet$, a vertical link, starting
from the lowest energy state. 
Placing the initial point of $\gamma_I$ at $(0,0)$, then
its final point is at $(M,N-M)$. The number of
up and right going paths from $(0,0)$ and $(N-M,M)$
is given by $C^N_M$, which equals the dimension
of $\h$.  This means that the path representation 
is faithful. 
In fig.~9 we show the paths associated to the 
ground state $I_0$ and an excited state $I$ at $g=0$. 
The path $\gamma_{I_0}$ is given by
$(0,0) \rightarrow (0,M) \rightarrow  (N-M,M)$, while 
the path associated to any excited state $I$ 
lies always below it. This fact makes possible to associate
a Young diagram $Y_I$ to every state $I$,
whose boundary is given by the paths
$\gamma_I$ and $\gamma_{I_0}$ with their
common links being removed (see fig.~9).
The GS is associated to the empty 
Young diagram, i.e. $Y_{I_0}$ = \O.
Using these definitions, the funtion
 $N_G(I)$  is given by the following algorithm
\cite{rsd}:
\begin{figure}[t!]
\begin{center}
\includegraphics[width= 9 cm,angle= 0]{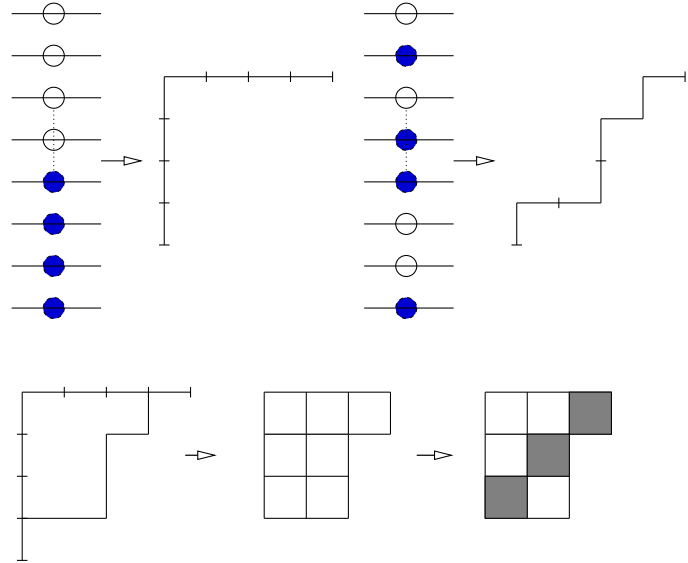}
\end{center}
\caption{Path representation of the ground state
$I_0$ and an excited state $I$. Combining 
these two paths one gets the Young diagram $Y_I$. 
The number of elementary 
excitations of the state $I$ is given by $N_G=3$, 
according to the rule (\ref{26}). 
}
\label{fig9}
\end{figure}
\beq
N_G(I) = {\rm number} \; {\rm of} \; 
{\rm boxes} \; {\rm on} \; {\rm the }\;
{\rm longest} \; {\rm SW-NE} 
 \;  {\rm diagonal }\; {\rm of }\; Y_I .
\label{26}
\eeq
In  the example shown in fig.9 we get
$N_G = 3$. The algorithm (\ref{26})
was proposed in reference \cite{rsd}
to explain a large amount of numerical simulations.
As explained in \cite{rsd} 
the physical mechanism underlying eq.~(\ref{26}) is 
the collective behavior of the holes and pairs
that occupy the closest levels to the Fermi level. 
There are also combinatorial reasons to support it (see below), 
however an analytical proof is needed.

\begin{figure}[h!]
\begin{center}
\includegraphics[width= 9.5 cm,angle= 0]{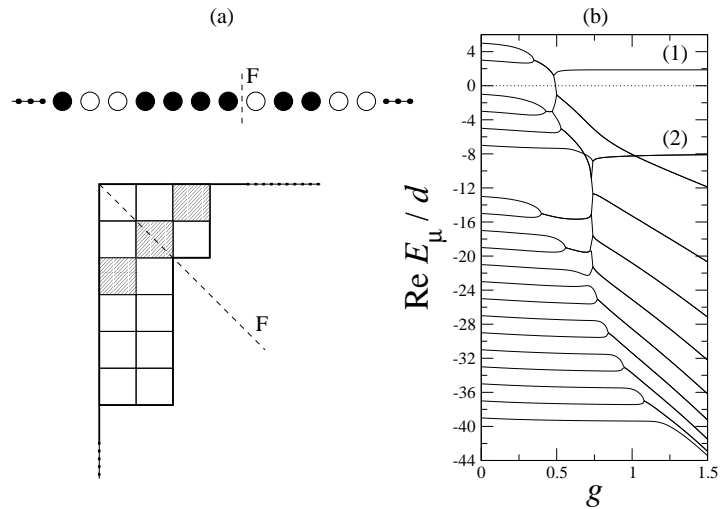}
\end{center}
\caption{a) The path and 
Young diagram of an excited state with $N_G=3$.  
The dotted line denotes the position of the Fermi level.
b) Real part of  $E_\mu$. For $g$ large enough  
there is a real root (1) and a complex
root (2) in agreement with the result $N_G=3$.}
\label{fig10}
\end{figure}

The non-triviality of (\ref{26}) can be 
seen in  fig.~10,
which shows the evolution
of the real part of the roots $E_\mu$ 
as a function of $g$. Only for sufficiently
large values of $g$, and after a complicated  pattern 
of fusion and splitting of roots
(2 real roots $\leftrightarrow$ 1 complex root) 
one observes that, indeed, $N_G=3$.
It seems very difficult to explain
this result on the basis of Bogoliubov's quasiparticle 
picture which 
only works properly for large number
of particles in the grand canonical ensemble.

We have  yet no proof of eq.~(\ref{26}), 
however there is the following consistency check
based on Gaudin's conjecture (\ref{23}):  
$d_{N,M,N_G}$ must be equal to the number of Young diagrams, 
$Y_I$, associated to 
the paths $\gamma_I$ going from $(0,0)$ to $(M,N-M)$, 
which have  $N_G$ boxes on their longest 
SW-NE diagonal.  The proof of this result uses the  methods of 
Ref.~\onlinecite{torres}. This result can also be proved
using the RSOS counting formulas.
(We thanks A.~Berkovich for pointing out this fact).  
As an example we show in fig.~11 how  
the paths corresponding to the case
$N=6, M=N_G=2$ can be mapped into RSOS paths. 

We would like finally to mention the
close relationship between the
path representation and the associated
Young diagrams of the BCS states with  
the crystal basis used in the Fock
space representation of the affine
quantum group $U_q( \widehat{ Sl(2)} )$ 
\cite{crystal,crystal1,crystal2}. 
At the moment of writting this paper we do not know
if there exists a quantum group structure underlying
the exactly solvable BCS model. This may indeed
be the case given that the integrability of the
BCS model can be explained from an inhomogenous
XXZ spin chain with boundary operators 
\cite{AFF,ZLMG,vDP,Scomo}.

\begin{figure}[h!]
\begin{center}
\includegraphics[width= 9.5 cm,angle= 0]{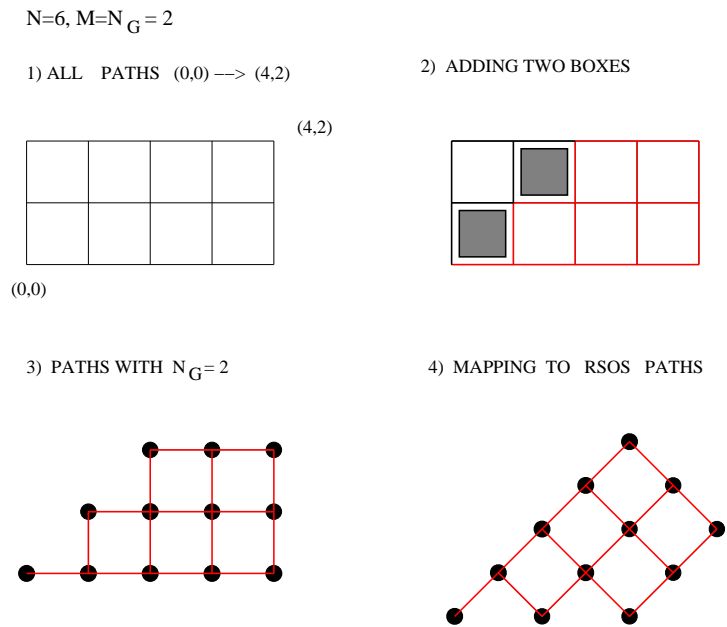}
\end{center}
\caption{1) Shows the $C^6_2 = 9$ paths
from $(0,0)$ to $(4,2)$, 2) The paths
corresponding to $N_G=2$ are those
below the  shaded boxes, 3) We select
only those paths with  $N_G=2$ and 4)
after a rotation the $N_G=2$ paths
become RSOS paths on a Bratelli diagram.}
\label{fig11}
\end{figure}

\section{Conclusions}

We have shown in reference \cite{rsd}
that the pair-hole excited states of 
the exactly solvable BCS model in the
canonical ensemble can be interpreted
in terms of elementary excitations
with peculiar counting properties
related to the Gaudin model, which
have no counterpart in the BCS-Bogoliubov's
picture of quasiparticles. 

The combinatorics involved in the counting
of elementary excitations is similar
to that of RSOS models, which suggests that
these excitations are in fact solitons
which were called ``gaudinos'' in \cite{rsd}.

Some open problems are 1) the relation between
these gaudinos and the Bogoliubov quasiparticles,
2) analytic proof of the algorithm for computing
the number of gaudinos $N_G(I)$ for each BCS 
state, 3) working out the physical consequences
of these excitations in the canonical ensemble
for small systems, 4) generalization of these
results to higher spin representations and
other Lie groups and supergroups.


\section*{Acknowledgments}

GS wants to thank A. Belavin and Y. Pugai for their kind
invitation to participate in the Chernogolovka workshop. 
This work has been supported by the grants 
BFM2000-1320-C02-01/02.

\end{document}